\documentclass[aps,prl,reprint,longbibliography,superscriptaddress]{revtex4-2}
\usepackage{mathrsfs}
\usepackage{amsmath,gensymb}
\usepackage{amsfonts}
\usepackage{amssymb}
\usepackage{amsthm}
\usepackage{graphicx}
\usepackage{natbib}
\usepackage{color}
\usepackage{bm}
\usepackage[caption=false]{subfig}
\usepackage{verbatim}
\usepackage[normalem]{ulem}
\usepackage{physics}
\usepackage{hyperref}

\begin{document}
\title{Ferron-Polaritons in Superconductor/Ferroelectric/Superconductor Heterostructures}

\author{M. Nursagatov}
\thanks{These authors contributed equally}
\affiliation{Moscow Institute of Physics and Technology, Dolgoprudny, 141700 Moscow region, Russia}

\author{Xiyin Ye}
\thanks{These authors contributed equally}
\affiliation{School of Physics, Huazhong University of Science and Technology, Wuhan 430074, China}

\author{G. A. Bobkov}
\affiliation{Moscow Institute of Physics and Technology, Dolgoprudny, 141700 Moscow region, Russia}

\author{Tao Yu}
\email{taoyuphy@hust.edu.cn}
\affiliation{School of Physics, Huazhong University of Science and Technology, Wuhan 430074, China}

\author{I. V. Bobkova}
\email{ivbobkova@mail.ru}
\affiliation{Moscow Institute of Physics and Technology, Dolgoprudny, 141700 Moscow region, Russia}

\begin{abstract} 
We predict the formation of ferron-polariton---a hybrid light-matter quasiparticle arising from the coupling between collective ferroelectric excitations (ferrons) and Swihart photons in a superconductor/ferroelectric/superconductor heterostructure. The coupling provides direct evidence for ferrons and reaches the ultrastrong-coupling regime, with a spectral gap in the terahertz range, orders of magnitude larger than those in magnetic analogues, reflecting the superior strength of electric dipole interactions. Our work establishes superconductor-ferroelectric heterostructures as a novel platform for exploring extreme light-matter coupling and for developing high-speed, ferroelectric-based quantum technologies at terahertz frequencies.

\end{abstract}

\maketitle

Ferroelectric materials, despite their ubiquity, have not been considered as active information conduits in microelectronics. Recent theoretical advances~\cite{Tang2022,Tang2024,Bauer2021,Bauer2022,Tang2022_thermoelectric,Bauer2023,Rodriguez-Suarez2024,Zhou2023,Zhu2024,Yu2026_review} have unveiled the existence of collective excitations of the ferroelectric order---polarization waves and their fundamental quanta, known as \textit{ferrons}. This discovery establishes an electric analogue to the well-known spin waves and their quanta magnons in magnetic systems~\cite{Kranendonk1958,Stancil_book,Chumak2015}. While coherent magnons are routinely observed and harnessed in a broad range of experiments~\cite{Kranendonk1958,Stancil_book,Chumak2015,Pirro2021,Kirilyuk2010,Nemec2018,Bae2022}, experimental studies of coherent ferrons remain scarce.

This disparity is especially striking, given that the inherent dipolar interaction between the electric dipoles vast exceeds their magnetic dipole counterparts. This can be quantified by the ratio of their
characteristic coupling energies: $\mathcal{E}_{\mathrm{FE}} / \mathcal{E}_{\mathrm{FM}} \sim (\alpha^{2} \epsilon_r)^{-1} \gg 1$, where $\alpha=1/137$ is the fine structure constant and $\epsilon_r$ is the high-frequency dielectric constant. A notable exception was recently demonstrated in the van der Waals ferroelectric materials $\mathrm{NbOI}_2$~\cite{Choe2025,Zhang2025}. Upon excitation by a short laser pulse, the generation and transport of polarization waves---that is, coherent ferrons---were observed. Their presence was detected via the emission of intense, narrow-band terahertz (THz) radiation at the ferroelectric transverse optical (TO) phonon frequency, and nonvolatile electric-field control of ferron excitations was demonstrated. Furthermore, efficient ferron injection and detection enabled by ferromagnetic metal contacts, achieving nonlocal signal transmission over micrometer distances, have been reported \cite{Shen2025}. Ferrons are predicted to mediate efficient electric polarization transport~\cite{Tang2022_thermoelectric,Bauer2021}, electric-field-tunable heat transport~\cite{Tang2022,Bauer2023,Wooten2023}, directional routing of information~\cite{Zhou2023}, and nonlocal thermoelectric effects~\cite{Tang2023}. Despite this recent progress, the field of ferronics is still hindered by the lack of direct experimental evidence for ferrons and, crucially, for their coupling to other excitations.

While the properties of bulk and surface ferrons have been explored theoretically~\cite{Tang2022,Tang2024,Bauer2022,Rodriguez-Suarez2024,Zhou2023,Yu2026_review}, their tunability and control are an open frontier.
Extending the analogy between magnonics and ferronics, hybrid quantum systems integrating superconductors with magnetic materials have emerged as a fertile platform for exploring strong light-matter interactions~\cite{Golovchanskiy2021,Golovchanskiy2021_phyrevapplied,Silaev2023,Qiu2024,Li2018,Golovchanskiy2020,Silaev2022,Gordeeva2025,Gordeeva2025_af,Yu2026_review,Popadiuk2026}. In superconductor/ferromagnet (S/F) heterostructures, for instance, ultrastrong coupling between magnons and confined microwave photons (Swihart modes) has been realized, leading to the formation of magnon-polaritons with nonclassical properties~\cite{Golovchanskiy2021,Golovchanskiy2021_phyrevapplied,Silaev2023,Qiu2024,Gordeeva2025,Yu2026_review}. Whether the superconductors can be used to tune ferron quantum properties with the potential for forming hybrid quasiparticles remains an open question.

\begin{figure}[htp!]
\centering
\includegraphics[width=0.98\linewidth]{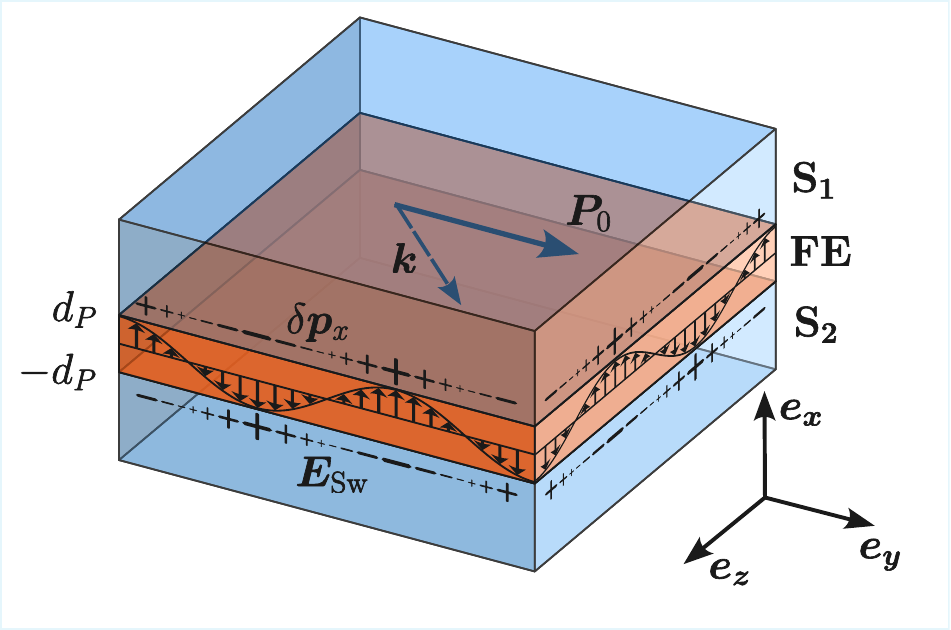}
\caption{Schematic of the S/FE/S heterostructure. The ferron mode polarized normal to the film interfaces, characterized by the polarization fluctuation $\delta \bm{p}_x$ (illustrated by the accompanying bound charges), couples to the in-plane-confined electric field $\bm{E}_{\rm Sw}$ of the Swihart photon mode in the superconducting resonator.}
\label{fig:sketch}
\end{figure}

In this work, we bridge this gap by investigating the collective electrodynamics of superconductor/ferroelectric/superconductor (S/FE/S) trilayers and demonstrate that they provide a natural and compelling counterpart to the magnonic platform. We show that in the planar geometry, as in Fig.~\ref{fig:sketch}, the ferron mode polarized normal to the film interfaces couples to the Swihart photon mode of the superconducting resonator~\cite{Swihart1961}, which forms \textit{ferron-polaritons}. 
This coupling is a direct consequence of the nonzero electric polarization of the excitation and would provide unambiguous experimental evidence for ferrons. Due to the extreme confinement of the electric field between the superconducting electrodes, this interaction reaches the \textit{ultrastrong coupling} regime. The characteristic energy scale of the interaction---the ferron-polariton gap---is orders of magnitude larger (on the scale of THz) than in magnetic analogues, reflecting the superior strength of electric dipole interactions. Our findings not only establish S/FE/S heterostructures as a novel platform for exploring ultrastrong light-matter coupling but also open a clear pathway toward ferroelectric-based quantum optics and signal processing at microwave and terahertz frequencies.

We consider a thin ferroelectric film of thickness $2d_P$ sandwiched between two semi-infinite superconducting layers, see Fig.~\ref{fig:sketch}. Assuming an uniaxial anisotropy with an easy axis along the $y$-direction in the plane of the ferroelectric film, the free energy density reads 
\begin{align}
    \mathcal{F} = \frac{\alpha_1}{2} P_y^2 + \frac{\alpha_2}{4} P_y^4 +\frac{\alpha_3} {2} (P_x^2 +P_z^2) - \frac{1}{2}{\bf E}_d \cdot \mathbf{P},
    \label{free_energy}
\end{align}
where the total electric polarization $\mathbf{P} = \bm P_0 + \delta \bm p = \{\delta p_x, P_{0y} + \delta p_y, \delta p_z\}$ is the sum of the static spontaneous polarization $\bm P_0$ and the fluctuation $\delta \bm p(\bm r, t)$, and $\alpha_{1,2,3}$ are the Landau parameters with $\alpha_1 <0$, $\alpha_2 > 0$, and $\alpha_3 >0$. In the linear-response regime $P_{0y} \gg \delta p_{x,y,z}$.
$P_{0y}=\pm \sqrt{-\alpha_1/\alpha_2}$ minimizes the free energy.
Typically, $P_0=\{0.746,0.753,0.265\}$~C/m$^{2}$ for LiNbO$_3$, PbTiO$_3$, and BaTiO$_3$~\cite{Tang2022}.

In the ferroelectrics, the electric polarization $\mathbf{P} = \Sigma_{i\in V} Q_i \mathbf{r}_i/V$ is governed by the ion charge $Q_i$ and ion position $\mathbf{r}_i$ of the $i$-th ion in a unit cell volume $V$. Under the effective electric field 
$\mathbf{E}_{\rm eff}=-\left({\partial F}/{\partial {\bf P}}\right)_T$,
where $F=\int d{\bf r} {\cal F}({\bf r})$ is the total free energy of the electric polarization, the dynamics of $\bm P$ follows Newton's Law $\overset{..}{\bf r} = (Q_j / m_j) \mathbf{E}_{\rm eff}$, where $m_j$ is the mass of the $j$-th ion, and  is thereby governed by the Landau-Khalatnikov-Tani (LKT) equation \cite{LKT1,LKT2,LKT3,LKT4,Yu2026_review} 
\begin{align}
    ({1}/\varepsilon _0 \Omega_p^2)\ddot{\bf P}={\bf E}_{\rm eff},
    \label{LKT}
\end{align}
where $\Omega_p^2={(\varepsilon_0 V)}^{-1} \sum_j {Q_j^2}/{m_j}$ is the  ionic plasma frequency.

Small fluctuations $\delta\mathbf{p}(\mathbf{r},t)$ around $\mathbf{P}_0$ are described by the linearized with respect to $\delta \bm p$ LKT equation
\begin{equation}
({1}/{\Omega_p^2})\delta\ddot{\mathbf{p}} + \hat{K}\,\delta\mathbf{p} = \varepsilon_0 \mathbf{E}_d,
\label{eq:LKT_linearized}
\end{equation}
where $\hat{K}=\mathrm{diag}(K_\perp,K_\parallel,K_\perp)$ with $K_\perp=\varepsilon_0\alpha_3$ and $K_\parallel=\varepsilon_0(\alpha_1+3\alpha_2 P_{0y}^2)$. The depolarization field $\mathbf{E}_d$ is induced by the polarization fluctuations themselves and is modified by the superconducting screens.

We focus on \textit{thin} ferroelectric films with the surface normal oriented along the $\hat{\bf x}$-direction.  
Since the film is thin, we assume the saturation electric polarization and the electric polarization fluctuation of frequency $\omega$ are uniform across the ferroelectric, i.e., 
\begin{align}
\bm P(\mathbf{r}, t)=(\bm P_0+\delta \tilde {\bm {p}}e^{i\mathbf{k} \cdot \pmb{\rho}-i\omega t})\left[
    \Theta(x+d_P)-\Theta(x-d_P)\right],
\label{P_full}
\end{align}
where \(\mathbf{r}=(x,\pmb{\rho})\), $\delta \tilde {\bm \rho} = (\delta \tilde{p}_{x}, \delta \tilde{p}_{y}, \delta \tilde{p}_{z})^T$ is the amplitude of the electric polarization fluctuation, $\bm k=(0,k_y,k_z)^T$ is the in-plane wave vector, and $\Theta(x)$ is the Heaviside step function. Mathematically, the thin-film limit means that the condition $k d_P \ll 1$ is fulfilled. 

We first address the dynamical electric ${\bf E}({\bf r},t)$ and magnetic ${\bf H}({\bf r},t)$ fields radiated by the ferronic excitations of the S/FE/S heterostructure. They obey Maxwell's equations 
\begin{eqnarray}
    \mathrm{rot}\;\bm{E} = -{\partial\bm{B}}/{\partial t}, ~~
    \mathrm{rot}\;\bm{H} = \bm{J} + {\partial\bm{D}}/{\partial t}, \label{eq:max}
\end{eqnarray}
which are supplemented by constitutive relations 
\begin{equation}
\bm{D} = \varepsilon_0\bm{E} + \bm{P},~~ 
    \bm{B} = \mu_0\bm{H},~~
    \bm{J} = \sigma\bm{E}.
    \label{eq:constitutive}
\end{equation}
Equations~(\ref{eq:max})-(\ref{eq:constitutive}) lead to the wave equation
\begin{equation}
\Delta\bm{H} = \sigma\mu_0\frac{\partial\bm{H}}{\partial t} + \varepsilon_0\mu_0\frac{\partial^2\bm{H}}{\partial t^2} -\mathrm{rot}\;\frac{\partial\bm{P}}{\partial t}.
\label{eq:wave}
\end{equation}
Taking the polarization in the form of Eq.~(\ref{P_full}) and seeking for the solution of Eq.~(\ref{eq:wave})  in the form $\bm{H}(\bm{r}, t) = \bm{\tilde H}(x)e^{i\bm{k\cdot\rho} - i\omega t}$ we obtain the following equations for the amplitude $\tilde {\bm H}(x)$ in the S and FE layers:
\begin{align}
&\text{in S}: ~~~    \frac{d^2\tilde{\bm H}}{d x^2} + (k_S^2-k^2) \tilde{\bm H} = 0, \label{eq:H_S}\\
    &\text{in FE}: ~    \frac{d^2\tilde{\bm H}}{d x^2} + (k_P^2-k^2) \tilde{\bm H} = i \omega \mathrm{rot} \bm P, 
\label{eq:H_FE}
\end{align}
where $k_S = \sqrt{\mu_0(i\omega\sigma_S + \omega^2\varepsilon_0)}$,   $k_P = \omega\sqrt{\varepsilon_0\mu_0}$, and $\sigma_S$ is the conductivity of the superconductor. Since the ferroelectric frequencies are in the terahertz range, the conductivity of the superconductor cannot be described by London's theory and needs to be calculated using the formalism of microscopic Green's functions. For isotropic BCS superconductors, which is the case under our consideration, we use the expression obtained in Ref.~\cite{ZIMMERMANN199199}
\begin{align}
    \sigma_S (\omega)=i\dfrac{\sigma_n}{2\omega\tau}I.
    \label{cond}
    \end{align}
Here $I$ is a dimensionless integral over quasiparticle energy, with its explicit expression given in Ref.~\cite{ZIMMERMANN199199}. Here we focus on the case $T \ll T_c$, where $T_c$ is the superconducting critical temperature, and $\hbar \omega < 2 \Delta$, where $\Delta$ is the superconducting order parameter. Then the real part of $\sigma_S$ is absent due to the absence of quasiparticles, and the conductivity Eq.~\eqref{cond} is well fitted by the expression
\begin{align}
    \sigma_S (\omega)=i/({\omega \mu_0 \lambda_{\rm{eff}}^2}), 
    \label{conductivity_lambda_eff}
\end{align}
which looks the same as that of London's theory \cite{Schmidt_book}, but the effective penetration depth $\lambda_{\rm{eff}}$ in general is not equal to London's penetration depth $\lambda_L = \sqrt{m_e/(\mu_0 \rho_s e^2)}$.

Substituting Eq.~(\ref{conductivity_lambda_eff}) into the expression for $k_S$ we obtain $k_S = \sqrt{\omega^2\varepsilon_0\mu_0 - 1/{\lambda_{\rm eff}^2}}$. Taking into account that at $\omega \sim 1$~THz and $\lambda_{\rm{eff}} \gtrsim 100$~nm, $k_S \approx i/\lambda_{\rm{eff}} \gg k$, where typical ferron-polariton wave numbers $k \sim \omega/c \sim 10^4 $~m/s. 

Equations~(\ref{eq:max})-(\ref{eq:constitutive}) allow to express the amplitude of the depolarization ferronic electric field $\bm E_d = \tilde {\bm E}(x) e^{i\bm{k\cdot\rho} - i\omega t}$ via $\tilde {\bm H}(x)$ and $\delta \tilde {\bm p}$. Taking into account that $k_S \gg k$ and neglecting terms of the order $(k/k_S)^2$ the resulting expressions read \cite{suppl}:
\begin{equation}
\begin{aligned}
    &\text{in S}:
    \left(\begin{matrix}
    \tilde E_y \\
    \tilde E_z \\
    \end{matrix}\right) = \frac{i\omega\mu_0}{k_S^2}\left(\begin{matrix}
    -\partial_x \tilde H_z \\
    \partial_x \tilde H_y \\
    \end{matrix}\right),\\
    &\text{in FE}:
    \left(\begin{matrix}
    \tilde E_y \\
    \tilde E_z \\
    \end{matrix}\right) = \frac{i\omega\mu_0}{k_{P}^2A^2}                              \left(\begin{matrix}
    \alpha & -\frac{K_1}{2} \\
    -\frac{K_1}{2} & \beta \\
    \end{matrix}\right)
    \left(\begin{matrix}
    -\partial_x \tilde H_z+i\omega \delta \tilde p_y \\
    \partial_x \tilde H_y+i\omega \delta \tilde p_z \\
    \end{matrix}\right),
\end{aligned}
\label{eq:E_resolved}
\end{equation}
where $A = \sqrt {k_P^2 - k^2}$, $\alpha = k_{P}^2-k_y^2$, $\beta = k_{P}^2-k_z^2$, and $K_1 = 2k_yk_z$.

Further solving Eqs.~(\ref{eq:H_S})-(\ref{eq:H_FE}) for $\tilde {\bm H}(x)$, expressing $\tilde {\bm E}(x)$ via $\tilde {\bm H}(x)$ and $\delta \tilde {\bm p}$ according to Eq.~(\ref{eq:E_resolved}) and making use of the boundary conditions, i.e., the continuity of $\{H_y, H_z, E_y, E_z\}$ at the S/FE interfaces $x=\pm d_P$, we arrive at the depolarization field at $|x|<d_P$ \cite{suppl}:
\begin{equation}
\mathbf{E}_d = -\hat{N}\,\delta\mathbf{p},
\label{eq:Edresponse}
\end{equation}
where, in the relevant thin-film limit $k d_P \ll 1$, $\tilde {\bm E}(x)$ can be considered uniform across the FE film and 
\begin{align}
&\hat{N} = \mathrm{diag}\big(N(k,\omega),0,0\big), \nonumber \\
& N(k,\omega)=\frac{\omega^2}{\varepsilon_0[\omega^2-c^2k^2 d_P/(d_P + \lambda_{{\rm eff}})]}.
\label{eq:Nmatrix}
\end{align}
The diagonal structure of $\hat{N}$ with the only nonzero element $N_x$ implies that only the fluctuation of the electric polarization $\delta p_x({\pmb \rho},t)$ along the film normal $\hat{\bf x}$-direction radiates the electric field along the $\hat{\bf x}$-direction in the ferroelectric film; while the fluctuations $\delta p_y({\pmb \rho},t)$ and $\delta p_z({\pmb \rho},t)$ generate negligible electric field.

At first, let us discuss the limit of vanishing screening, $\lambda_{\mathrm{eff}} \to \infty$. This limit is somewhat artificial: it does not correspond to normal metal leads, as we neglect the real part of $\sigma_S$, which is inevitably present in a normal metal due to quasiparticles. Nor does it correspond to a thin FE film in vacuum, since it does not allow for an oscillating electromagnetic field outside the FE film. Nevertheless, this limit is instructive because it corresponds to vanishing coupling between the ferron and the Swihart photon---the Swihart frequency 
\begin{equation}
\Omega_s(k) = ck \sqrt{\frac{d_P}{d_P+\lambda_{\rm eff}}},
\label{eq:Swihart}
\end{equation} tends to zero in this limit (see also the quantum derivation of the coupling constant below). 

The depolarization field in this limit is obtained from Eqs.~\eqref{eq:Edresponse}--\eqref{eq:Nmatrix} by taking $\lambda_{\mathrm{eff}} \to \infty$, yielding
\begin{align}
    E_{d,x} \approx -{\delta p_{x}(\bm{\rho},t)}/{\varepsilon_0}, \qquad
    E_{d,y,z} \to 0,
    \label{eq:Ed_vacuum}
\end{align}
inside the FE film. Substituting Eq.~\eqref{eq:Ed_vacuum} into the linearized LKT equation~\eqref{eq:LKT_linearized} gives the ferron eigenfrequencies, which separate into three branches with the following polarization directions:
\begin{align}
    \omega_{1}  &= \Omega_p \sqrt{1+K_{\perp}}, \quad \mathbf{e}_{1} = (1,0,0)^T, \nonumber\\
    \omega_{+}  &= \Omega_p \sqrt{K_{\parallel}}, \quad \mathbf{e}_{+} = (0,1,0)^T, \nonumber\\
    \omega_{-}  &= \Omega_p \sqrt{K_{\perp}}, \quad \mathbf{e}_{-} = (0,0,1)^T.
    \label{eq:FE_vacuum_frequencies}
\end{align}

For LiNbO$_3$ at room temperature, the Landau parameters $\alpha_1 = -2.012 \times 10^{9}~\mathrm{N\cdot m^2/C^2}$, $\alpha_2 = 3.608 \times 10^{9}~\mathrm{N\cdot m^6/C^4}$, and $\alpha_3 = 1.345 \times 10^{9}~\mathrm{N\cdot m^2/C^2}$ \cite{Scrymgeour2005}, with an ionic plasma frequency $\Omega_p = 6.39~\mathrm{THz}$ \cite{Rodriguez-Suarez2024}. These values yield the dimensionless stiffness coefficients $K_{\perp} = 0.012$ and $K_{\parallel} = 0.036$, a zero-field static polarization $P_0 = 0.746~\mathrm{C/m^2}$, and the ferron frequencies $\omega_1 = 6.43~\mathrm{THz}$, $\omega_+ = 1.21~\mathrm{THz}$, and $\omega_- = 0.7~\mathrm{THz}$.

Taking into account the superconducting leads and substituting Eqs.~(\ref{eq:Edresponse})-(\ref{eq:Nmatrix}) into (\ref{eq:LKT_linearized}), we obtain the modified equations for the ferron-polariton eigenfrequencies
\begin{align}
    &\omega_{S,1}^2=\Omega_p^2 (\varepsilon_0 N(k,\omega)+K_{\perp}), \nonumber\\
    &\omega_{S,+} =\omega_{+} = {\Omega_p}\sqrt{K_{\parallel}}, \nonumber\\
    &\omega_{S,-}=\omega_{-} = \Omega_p\sqrt{K_{\perp}}.
    \label{eq:S_FE_S_frequencies}
\end{align}

\begin{figure}
    \centering
    \includegraphics[width=0.95\linewidth]{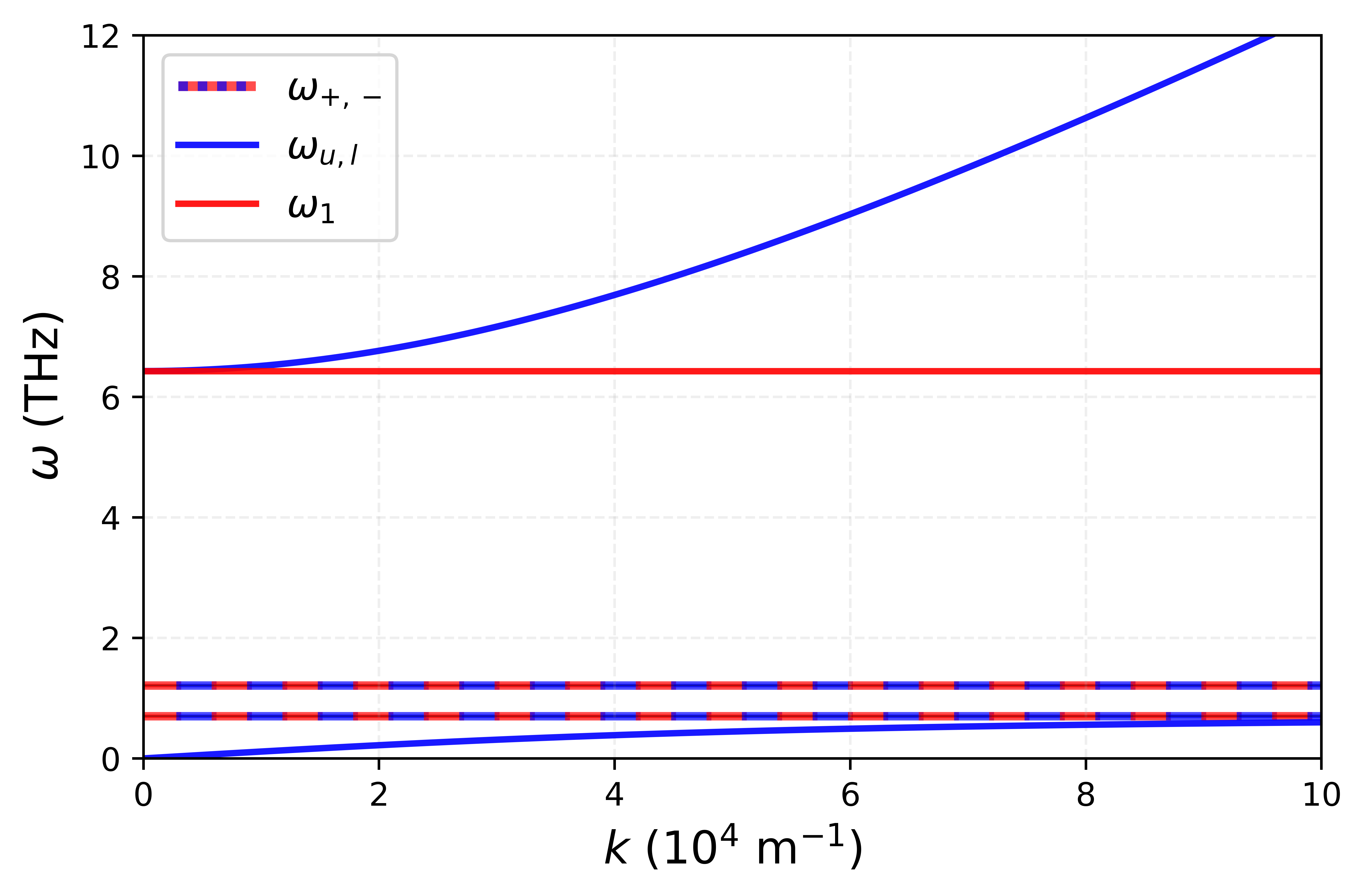}
    \caption{Spectra of ferron-polariton and ferron excitations of the S/FE/S structure. $d_P/\lambda_{\rm eff}=1/8$, $\{\alpha_1,\alpha_2,\alpha_3\} = \{-2.012,3.608,1.345\}\times 10^9 \; \text{Nm/C}^2$, and $\Omega_p=6.39$ THz. The $\delta p_x$-ferron-polariton branches $\omega_{u,l}$ are shown by the blue curves. The $\delta p_x$-ferron dispersion $\omega_1$ in the limit $\lambda_{\rm eff} \to \infty$ is shown by the red curve. $\delta p_{y,z}$-ferron frequencies $\omega_\pm$, which are the same in the presence of the superconducting screening and at $\lambda_{\rm eff} \to \infty$, are shown by blue-red dashed curves.}
    \label{fig:placeholder}
\end{figure}

According to Eq.~\eqref{eq:S_FE_S_frequencies}, the eigenfrequencies $\omega_{S,\pm}$ for $\delta p_{y,z}$ ferron excitations remain unchanged by the superconducting screening. As implied by Eq.~\eqref{eq:Edresponse}, these excitations induce no depolarization field. In contrast, a $\delta p_x$ excitation generates a depolarization field. However, within the S/FE/S heterostructure, screening by the superconductors confines this field to a scale of $\sim 2(d + \lambda_{\rm eff})$. This confinement enables strong, efficient coupling to the electric field of the superconducting Swihart photon, which hybridizes with the ferron mode, leading to the formation of a ferron-polariton.

Substitution of $N(k,\omega)$ from Eq.~(\ref{eq:Nmatrix}) into (\ref{eq:S_FE_S_frequencies}) yields the upper and lower ferron-polariton branches
\begin{align}
\omega_{u,l}^2 = \frac{1}{2}\left( \Omega_s^2 + \omega_1^2 \pm \sqrt{(\omega_1^2-\Omega_s^2)^2+4 \Omega_s^2 \Omega_p^2} \right).   
\label{eq:ferron_polaritons_classical}
\end{align}

From a quantum view, the Swihart photon in the S/FE/S resonator at $\bm P = 0$ has a frequency \eqref{eq:Swihart}. The magnetic field $\hat{\bf H}_{Sw}=\hat{H}_{{\rm Sw},y}\hat{\bf y}+\hat{H}_{{\rm Sw},z}\hat{\bf z}$ inside the insulator in terms of the photon operators $\hat p_{\bm k}$ is quantized as \cite{Qiu2024}
\begin{align}
     \hat{H}_{{\rm Sw},y}&=\sum_{\bf k} \left(\frac{\Omega_s({\bf k})}{2k}\sqrt{\frac{\epsilon_0\hbar\Omega_s({\bf k})}{d_P S}}\cos\theta_{\bf k} e^{i{\bf k}\cdot{\pmb \rho}}\hat{p}_{\bf k}+{\rm H.c.}\right),\nonumber\\
     \hat{H}_{{\rm Sw},z}&=\sum_{\bf k} \left(-\frac{\Omega_s({\bf k})}{2k}\sqrt{\frac{\epsilon_0\hbar\Omega_s({\bf k})}{d_P S}}\sin\theta_{\bf k} e^{i{\bf k}\cdot{\pmb \rho}}\hat{p}_{\bf k}+{\rm H.c.}\right),
     \label{magnetic_field_quantization2}
\end{align}
where $\theta_{\bf k}$ is the angle between the wave vector ${\bm k}$ and the $\hat{\bf z}$-direction and $S$ is the area of the S/FE interface. They are uniform across the insulator film. According to 
$\nabla\times {\bf H}=\partial_t{\bf D}=\epsilon_0\partial_t{\bf E}$, inside the interlayer the electric field~\cite{suppl}
\begin{align}
    {\bf E}_{\rm Sw}({\bf r})=\sum_{\bf k} \frac{1}{2\epsilon_0}\sqrt{\frac{\epsilon_0\hbar\Omega_s({\bf k})}{d_P S}}e^{i{\bf k}\cdot{\pmb \rho}}\left(\hat{p}_{\bf k}+\hat{p}^{\dagger}_{-\bf k}\right)\hat{\bf x}
    \label{eq:swihart_quantum}
\end{align}
is along the film normal $\hat{\bf x}$-direction.

On the other hand, the quantized expression for the polarization fluctuation reads \cite{suppl}
\begin{align}
    \begin{pmatrix}
        \delta \hat{p}_{x}(\bf{r}) \\
        \delta \hat{p}_{y}(\bf{r})\\
        \delta \hat{p}_{z}(\bf{r})
    \end{pmatrix} &=\frac{1}{\sqrt{V_P}}\sum_{\lambda,{\bf k}}\sqrt{\frac{\hbar}{2m_p\omega_{\lambda}}}{\bf e}_{\lambda} (\hat{a}_{\lambda,{\bf k}}+\hat{a}^{\dagger}_{\lambda,-{\bf k}}) \nonumber \\
    &\times e^{i{\bf k}\cdot \pmb{\rho}}\left[\Theta(x+d_P)-\Theta(x-d_P)\right],
    \label{delta_p_a}
\end{align}
where $\hat a_{\lambda, \bm k}$ is the ferron operator, the eigenfrequency and the polarization direction $\bm e_{\lambda}$ of the mode $\lambda=\{1,+,-\}$ are expressed by Eq.~(\ref{eq:FE_vacuum_frequencies}), and $V_P$ is a volume of the ferroelectric insulator. 

The polarization fluctuation \eqref{delta_p_a} and electric field \eqref{eq:swihart_quantum} couples according to 
\begin{align}
    \hat{H}_{\rm int}&=-\int dxd{\pmb \rho}\hat{\bf P}({\bf r})\cdot\hat{\bf E}_{\rm Sw}({\bf r})\nonumber\\
    &=\sum_{\bf k}\hbar g({\bf k})(\hat{a}_{1,{\bf k}}+\hat{a}_{1,-{\bf k}}^{\dagger})(\hat{p}_{-{\bf k}}+\hat p^{\dagger}_{\bf k}),
    \label{eq:g}
\end{align}
 where the coupling constant 
\begin{align}
g(k)=-({\Omega_p}/{2})\sqrt{{\Omega_s(k)}/{\omega_{1}}}
\end{align}
is isotropic.  In full agreement with the classical calculation, only the $\delta p_x$-ferron (mode ``1") couples to the Swihart photon because its electric field has only surface-normal $x$-component. 

The full quantum Hamiltonian of the S/FE/S heterostructure
\begin{align}
    \hat{H} &=\sum_{\lambda, {\bf k}}\hbar\omega_{\lambda}(\hat{a}^{\dagger}_{\lambda,{\bf k}}\hat{a}_{\lambda,{\bf k}}+{1}/{2}) + \sum_{\bf k}\hbar\Omega_s(k)(\hat{p}_{\bf k}^\dagger\hat{p}_{\bf k}+{1}/{2})  \nonumber \\
    &+\sum_{\bf k}\hbar g(k)(\hat{a}_{1,{\bf k}}+\hat{a}_{1,-{\bf k}}^{\dagger})(\hat{p}_{-{\bf k}}+\hat p^{\dagger}_{\bf k}).
\label{eq:ham_quantum}    
\end{align}
The Hamiltonian is diagonalized in terms of ferron-polariton operators
\begin{align}
    \hat d_{j,\bm k} = a_{j,\bm k} \hat a_{1,\bm k} + p_j \hat p_{\bm k} + \tilde a_{j,\bm k}\hat a_{1,-\bm k}^\dagger + \tilde p_j \hat p_{-\bm k}^\dagger,
\label{eq:FP}    
\end{align}
where $j=\{u,l\}$ is the index of upper and lower ferron-polariton branches. The resulting eigenfrequencies of the ferron-polariton quasiparticles are given by Eq.~(\ref{eq:ferron_polaritons_classical}). At resonance, $\Omega_s({\bf k})=\omega_{1}$, $|g({\bf k})|=\Omega_p/2$. 

The dispersion relations of the resulting ferron-polariton modes are plotted in Fig.~\ref{fig:placeholder}. The $\delta p_x$-ferron-polariton branches $\omega_{u,l}$ are shown by the blue curves, which differ from the $\delta p_x$-ferron dispersion $\omega_1$ in the limit $\lambda_{\rm eff} \to \infty$ shown by the red curve. $\delta p_{y,z}$-ferron frequencies, which are the same in the presence of the superconducting screening and at $\lambda_{\rm eff} \to \infty$, are shown by blue-red dashed curves. 

Our treatment is applicable in the frequency region $\omega < 2\Delta(T=0)/\hbar$; otherwise, the excitation of quasiparticles by the electromagnetic field should be taken into account even at $T=0$, which leads to a finite real part of the S layer conductivity \cite{Abrikosov1958,Abrikosov1960,Mattis1958,ZIMMERMANN199199}. Taking Nb with $T_c \approx 9$~K as a superconductor, this condition gives that the theory is applicable at $\omega< 4$~THz, and for NbN with $T_c \approx 15$~K---at $\omega< 6.9$~THz. Nonetheless, we present a wide frequency range to elucidate and discuss the fundamental physics of ferron-polaritons in S/FE/S heterostructures. Notably, the coupling between the $\delta p_x$-ferron and the photon appears to be far in the ultra-strong coupling regime in that the anticrossing gap $\Delta \omega = \omega_1 (\sqrt{1+\Omega_p/\omega_1}-\sqrt{1-\Omega_p/\omega_1})$ between the lower and upper ferron-polariton branches is very close to the bare ferron frequency $\omega_1$ itself. 

In contrast to the S/F/S and S/AF/S heterostructures—where the formation of magnon-polaritons via ultrastrong magnon–Swihart-photon coupling has also been predicted—the predicted spectral gap is markedly larger in S/FE/S systems. For S/F/S systems, the gap amounts to a few GHz \cite{Silaev2023,Qiu2024,Gordeeva2025}, while for S/AF/S systems it has been predicted to reach about 100 GHz \cite{Gordeeva2025_af}. Our calculation for S/FE/S heterostructures, however, predicts a substantially larger gap on the order of several THz. Physically, the magnitude of this gap is set by the characteristic energy of the dipolar interaction between the polarization $\delta \bm p$ and the depolarization electric field $\bm E_d$, which significantly exceeds the dipolar energy between the magnetic moment and demagnetization field.

Besides, unlike their magnon-polariton counterparts in $\mathrm{S/F/S}$~\cite{Qiu2024,Gordeeva2025} and $\mathrm{S/AF/S}$~\cite{Gordeeva2025_af} systems, the ferron-polaritons in $\mathrm{S/FE/S}$ heterostructures are isotropic. In magnetic systems, the magnon–Swihart-photon coupling is strongly anisotropic, being maximal for $\bm{k} \parallel \bm{M}_0$ and vanishing for $\bm{k} \perp \bm{M}_0$~\cite{Qiu2024}, where $\bm{M}_0$ is the equilibrium magnetization. In contrast, for $\mathrm{S/FE/S}$ heterostructures, the ferron–photon interaction shows no angular dependence on the direction of the equilibrium polarization $\bm{P}_0$. This isotropy originates from the distinct coupling mechanisms: magnon-polaritons form due to the coupling between the magnon and the in-plane \textit{magnetic} field of the Swihart mode (whose direction is perpendicular to $\bm{k}$). Conversely, ferron-polaritons arise from the coupling between the ferron and the out-of-plane \textit{electric} field of the same mode, whose orientation is independent of $\bm{k}$.

In summary, we predict the formation of ferron-polaritons—hybrid light-matter quasiparticles arising from the coupling between collective ferroelectric excitations (ferrons) and Swihart photons in a superconductor/ferroelectric/superconductor heterostructure. The coupling, mediated by the out-of-plane electric field of the Swihart mode, provides unambiguous experimental evidence for ferrons. It is highly selective: only ferron modes polarized normal to the film interact, while in-plane modes remain dark.  Crucially, the interaction reaches the ultrastrong coupling regime with a spectral gap in the THz range, orders of magnitude larger than in analogous magnetic (S/F/S and S/AF/S) systems, reflecting the fundamental superiority of electric over magnetic dipole energy scales. The resulting ferron-polariton spectrum is fundamentally isotropic, in stark contrast to its anisotropic magnonic counterparts. 

Our work establishes S/FE/S heterostructures as a novel and potent platform for exploring extreme light-matter interactions at terahertz frequencies. It bridges the fields of ferroelectricity, superconductivity, and quantum optics, opening pathways for developing high-speed, low-loss signal processing devices and quantum technologies based on ferroelectric excitations.

\begin{acknowledgments}
{\it Acknowledgments.}---We thank A. M. Bobkov for stimulating and fruitful discussions. The financial support by Grant from the ministry of science and higher education of the Russian Federation No. 075-15-2025-010 (dipolar fields calculations), from the Russian
Science Foundation via the project No.~23-72-30004 (calculations of ferron-polariton dispersions), and from the National Key Research and Development Program of China under Grant No.~2023YFA1406600 and the National Natural Science Foundation of China under Grant No.~12374109 is acknowledged.   
\end{acknowledgments}

\begin{widetext}

\section*{Supplemental Material for the Letter \\
	 ``Ferron-Polaritons in Superconductor/Ferroelectric/Superconductor Heterostructures''}

\section{Depolarization Electric Field}

\subsection{Maxwell's Equations}

We begin with Maxwell's equations
\begin{subequations}
\begin{align}
    \curl \mathbf{E} &= - \partial_t \mathbf{B}, \label{eq:maxwell1}\\
    \curl \mathbf{H} &= \mathbf{J} + \partial_t \mathbf{D}, \label{eq:maxwell2}
\end{align}
\end{subequations}
incorporating a polarization source term $\mathbf{P}$ in the ferroelectrics
with the constitutive relations
\begin{subequations}
\begin{align}
    \mathbf{D} &= \varepsilon_0 \mathbf{E} + \mathbf{P},  \label{eq:cons1} \\
    \mathbf{B} &= \mu_0 \mathbf{H},  \label{eq:cons2} \\
    \mathbf{J} &= \sigma \mathbf{E}.
    \label{eq:cons3}
\end{align}
\end{subequations}

To derive a wave equation for the magnetic field $\mathbf{H}$, we take the curl of Eq.~\eqref{eq:maxwell2} and substitute Eqs.~\eqref{eq:maxwell1} and \eqref{eq:cons1}-\eqref{eq:cons3}:
\begin{align}
    \curl \curl \mathbf{H} &= \sigma \curl \mathbf{E} + \varepsilon_0 \partial_t \curl \mathbf{E} + \curl \partial_t \mathbf{P} \nonumber \\
    &= -\sigma \mu_0 \partial_t \mathbf{H} - \varepsilon_0 \mu_0 \partial_t^2 \mathbf{H} + \curl \partial_t \mathbf{P}.
\end{align}
Using the vector identity $\curl \curl \mathbf{H} = \grad (\div \mathbf{H}) - \nabla^2 \mathbf{H}$ and noting that $\div \mathbf{B} = \mu_0 \div \mathbf{H} = 0$, we obtain
\begin{equation}
    \nabla^2 \mathbf{H} = \sigma \mu_0 \partial_t \mathbf{H} + \varepsilon_0 \mu_0 \partial_t^2 \mathbf{H} - \curl \partial_t \mathbf{P}. \label{eq:wave_eq_H}
\end{equation}

\subsection{Plane-Wave Ansatz and Field Solutions}
We model the polarization as a static spontaneous polarization plus a small, propagating fluctuation:
\begin{equation}
    \mathbf{P}(\mathbf{r}, t) = \mathbf{P}_0 + \delta \tilde{\mathbf{p}} e^{i \mathbf{k} \cdot \bm{\rho} - i \omega t},
\end{equation}
where $\mathbf{r} = (x, \bm{\rho})$, $\bm{\rho} = (y, z)$, and $\mathbf{k} = (k_y, k_z)$ is the in-plane wave vector.

We seek solutions for the magnetic field of the form
\begin{equation}
    \mathbf{H}(\mathbf{r}, t) = \tilde{\mathbf{H}}(x) e^{i \mathbf{k} \cdot \bm{\rho} - i \omega t}.
\end{equation}
Substitution of this ansatz into Eq.~\eqref{eq:wave_eq_H} yields the following equations for the amplitude $\tilde{\mathbf{H}}(x)$ in the superconducting (S) and ferroelectric (FE) layers:
\begin{subequations}
\begin{align}
    \text{(S layers)}&: \quad \partial_x^2 \tilde{\mathbf{H}} + \qty(k_S^2 - k^2) \tilde{\mathbf{H}} = 0, \label{eq:H_S_supp} \\
    \text{(FE layer)}&: \quad \partial_x^2 \tilde{\mathbf{H}} + \qty(k_P^2 - k^2) \tilde{\mathbf{H}} = -\omega \bm k \times \delta \tilde{\mathbf{p}}. \label{eq:H_P}
\end{align}
\end{subequations}
Here, $k^2 = k_y^2 + k_z^2$, $k_S = \sqrt{i \omega \sigma_S \mu_0 + \omega^2 \varepsilon_0 \mu_0}$, and $k_P = \omega \sqrt{\varepsilon_0 \mu_0}$. For a superconductor at $T \ll T_c$ and $\hbar\omega < 2\Delta$, the conductivity is purely imaginary and can be approximated as $\sigma_S(\omega) \approx i / (\omega \mu_0 \lambda_{\text{eff}}^2)$, leading to $k_S \approx \sqrt{\omega^2 \varepsilon_0 \mu_0 - 1/\lambda_{\text{eff}}^2} \approx i/\lambda_{\text{eff}}$ for typical THz frequencies.

Solving Eqs.~\eqref{eq:H_S_supp}-\eqref{eq:H_P} in each layer we obtain
\begin{equation}
\begin{aligned}
    &\text{in S1: }(x>d_P)\;\;\; \bm{\tilde H} = \bm S_1 e^{iB(x-d)},\\
    &\text{in S2: }(x<d_P)\;\;\; \bm{\tilde H} = \bm S_2 e^{-iB(x+d)},\\
    &\text{in FE: }(|x|<d_P)\;\;\; \bm{\tilde H} = \bm C_1 e^{iAx} + \bm C_2 e^{-iAx} - \frac{\omega}{A^2}\bm k\times \delta \tilde{\bm p},
    \label{eq:H_solution}
\end{aligned}
\end{equation}
where
$$
A = \sqrt{k_P^2 - k^2}, \;\;\; B = \sqrt{k_S^2 - k^2}.
$$

From Maxwell's equations, the electric field amplitude $\tilde{\mathbf{E}}(x)$ can be expressed in terms of $\tilde{\mathbf{H}}(x)$ and $\tilde{\mathbf{P}}$. The general relation from Eq.~\eqref{eq:maxwell2} is
\begin{equation}
    \mathbf{E}_d = \frac{\curl \mathbf{H} + i\omega \delta \mathbf{p}}{\sigma - i\omega \varepsilon_0}. \label{eq:E_general}
\end{equation}
The explicit components of the electric field follow directly from Eq.~(\ref{eq:E_general}):
\begin{equation}
    \begin{aligned}
        \tilde{E}_x &= \frac{i k_y \tilde{H}_z - i k_z \tilde{H}_y + i \omega \delta \tilde{p}_x}{\sigma - i \omega \varepsilon_0}, \\[2pt]
        \tilde{E}_y &= \frac{i k_z \tilde{H}_x - \partial_x \tilde{H}_z + i \omega \delta \tilde{p}_y}{\sigma - i \omega \varepsilon_0}, \\[2pt]
        \tilde{E}_z &= \frac{\partial_x \tilde{H}_y - i k_y \tilde{H}_x + i \omega \delta \tilde{p}_z}{\sigma - i \omega \varepsilon_0}.
    \end{aligned}
    \label{eq:E_components}
\end{equation}
The $x$-component of the magnetic field can be obtained from Eq.~(\ref{eq:maxwell1}):

\begin{equation}
    \tilde{H}_x = \frac{i k_y \tilde{E}_z - i k_z \tilde{E}_y}{i \omega \mu_0}.
    \label{eq:Hx_from_E}
\end{equation}

Substitution of Eq.~\eqref{eq:Hx_from_E} into the in-plane electric field components $\tilde{E}_y$ and $\tilde{E}_z$, we can express them via the in-plane magnetic field components $\tilde{H}_y$ and $\tilde{H}_z$:
\begin{equation}
    \begin{pmatrix}
        k_{S/P}^2 - k_z^2 & k_z k_y \\
        k_z k_y & k_{S/P}^2 - k_y^2
    \end{pmatrix}
    \begin{pmatrix}
        \tilde{E}_y \\
        \tilde{E}_z
    \end{pmatrix}
    = i \omega \mu_0
    \begin{pmatrix}
        -\partial_x \tilde{H}_z + i \omega \delta \tilde{p}_y \\
        \partial_x \tilde{H}_y + i \omega \delta \tilde{p}_z
    \end{pmatrix}.
    \label{eq:E_matrix_form}
\end{equation}
Solving for the electric field components, we obtain
\begin{equation}
        \begin{pmatrix}
        \tilde{E}_y \\
        \tilde{E}_z
    \end{pmatrix}
    = \frac{i \omega \mu_0}{k_{S/P}^2 (k_{S/P}^2 - k^2)}
    \begin{pmatrix}
        k_{S/P}^2 - k_y^2 & -k_z k_y \\
        -k_z k_y & k_{S/P}^2 - k_z^2
    \end{pmatrix}
    \begin{pmatrix}
        -\partial_x \tilde{H}_z + i \omega \tilde{P}_y \\
        \partial_x \tilde{H}_y + i \omega \tilde{P}_z
    \end{pmatrix}.
    \label{eq:E_solution_general}
\end{equation}

In the superconducting electrodes, where the wave number satisfies $k_S \gg k$, we can simplify this expression by neglecting terms of order $(k/k_S)^2$:
\begin{equation}
    \text{in S:} \quad
    \begin{pmatrix}
        \tilde{E}_y \\
        \tilde{E}_z
    \end{pmatrix}
    = \frac{i \omega \mu_0}{k_S^2}
    \begin{pmatrix}
        -\partial_x \tilde{H}_z \\
        \partial_x \tilde{H}_y
    \end{pmatrix}.
    \label{eq:E_in_S}
\end{equation}

Within the ferroelectric (P) layer, we retain the full form but express it using the parameters characteristic of the dielectric:

\begin{equation}
    \text{in FE:} \quad
    \begin{pmatrix}
        \tilde{E}_y \\
        \tilde{E}_z
    \end{pmatrix}
    = \frac{i \omega \mu_0}{k_P^2 A^2}
    \begin{pmatrix}
        \alpha & -K_1/2 \\
        -K_1/2 & \beta
    \end{pmatrix}
    \begin{pmatrix}
        -\partial_x \tilde{H}_z + i \omega \delta \tilde{p}_y \\
        \partial_x \tilde{H}_y + i \omega \delta \tilde{p}_z
    \end{pmatrix},
    \label{eq:E_in_P}
\end{equation}
where we have defined
\begin{equation}
    \alpha = k_P^2 - k_y^2, \quad 
    \beta = k_P^2 - k_z^2, \quad 
    K_1 = 2 k_y k_z.
    \label{eq:def_parameters}
\end{equation}
Equations (\ref{eq:E_in_S}) and (\ref{eq:E_in_P}) provide the resolved expressions for the in-plane electric field components necessary for applying the electromagnetic boundary conditions at the S/FE interfaces.

\subsection{Boundary Conditions at the S/FE Interfaces}
The electromagnetic boundary conditions require the continuity of the tangential field components $H_y$, $H_z$, $E_y$, and $E_z$ at the interfaces $x = \pm d_P$. Applying these conditions to the general solutions given by Eq.~\eqref{eq:H_solution} yields the following system of equations.

For the $y$-component of the magnetic field,
\begin{subequations}
\begin{align}
    S_{1y} &= C_{1y} e^{iAd_P} + C_{2y} e^{-iAd_P} - \frac{\omega}{A^2} k_z \delta \tilde{p}_x, \label{eq:bc_Hy_plus} \\
    S_{2y} &= C_{1y} e^{-iAd_P} + C_{2y} e^{iAd_P} - \frac{\omega}{A^2} k_z \delta \tilde{p}_x. \label{eq:bc_Hy_minus}
\end{align}
\end{subequations}
For the $z$-component of the magnetic field,
\begin{subequations}
\begin{align}
    S_{1z} &= C_{1z} e^{iAd_P} + C_{2z} e^{-iAd_P} + \frac{\omega}{A^2} k_y \delta \tilde{p}_x, \label{eq:bc_Hz_plus} \\
    S_{2z} &= C_{1z} e^{-iAd_P} + C_{2z} e^{iAd_P} + \frac{\omega}{A^2} k_y \delta \tilde{p}_x. \label{eq:bc_Hz_minus}
\end{align}
\end{subequations}
The continuity of the in-plane electric field components, using Eqs.~(\ref{eq:E_in_S}) and (\ref{eq:E_in_P}), gives
\begin{subequations}
\begin{align}
    \frac{1}{k_S} S_{1z} &= \frac{1}{k_P^2 A^2} \Bigg[ \alpha \Big( A (C_{1z} e^{iAd_P} - C_{2z} e^{-iAd_P}) - \omega \delta \tilde{p}_y \Big) + \frac{K_1}{2} \Big( A (C_{1y} e^{iAd_P} - C_{2y} e^{-iAd_P}) + \omega \delta \tilde{p}_z \Big) \Bigg], \label{eq:bc_Ey_plus} \\
    \frac{1}{k_S} S_{1y} &= \frac{1}{k_P^2 A^2} \Bigg[ \frac{K_1}{2} \Big( A (C_{1z} e^{iAd_P} - C_{2z} e^{-iAd_P}) - \omega \delta \tilde{p}_y \Big) + \beta \Big( A (C_{1y} e^{iAd_P} - C_{2y} e^{-iAd_P}) + \omega \delta \tilde{p}_z \Big) \Bigg], \label{eq:bc_Ez_plus} \\
    -\frac{1}{k_S} S_{2z} &= \frac{1}{k_P^2 A^2} \Bigg[ \alpha \Big( A (C_{1z} e^{-iAd_P} - C_{2z} e^{iAd_P}) - \omega \delta \tilde{p}_y \Big) + \frac{K_1}{2} \Big( A (C_{1y} e^{-iAd_P} - C_{2y} e^{iAd_P}) + \omega \delta \tilde{p}_z \Big) \Bigg], \label{eq:bc_Ey_minus} \\
    -\frac{1}{k_S} S_{2y} &= \frac{1}{k_P^2 A^2} \Bigg[ \frac{K_1}{2} \Big( A (C_{1z} e^{-iAd_P} - C_{2z} e^{iAd_P}) - \omega \delta \tilde{p}_y \Big) + \beta \Big( A (C_{1y} e^{-iAd_P} - C_{2y} e^{iAd_P}) + \omega \delta \tilde{p}_z \Big) \Bigg]. \label{eq:bc_Ez_minus}
\end{align}
\end{subequations}

\subsection{Thin-Film Approximation}
We now consider the physically relevant limit of a \textit{thin} ferroelectric film, where the phase accumulation across it is small, i.e., $A d_P \ll 1$. In this limit, we approximate the exponentials as $e^{\pm i A d_P} \approx 1 \pm i A d_P$. To simplify the resulting algebraic system, it is convenient to introduce symmetric ($s$) and antisymmetric ($a$) combinations of the magnetic field coefficients within the ferroelectric layer:
\begin{equation}
    C_y^s = C_{1y} + C_{2y}, \quad C_y^a = C_{1y} - C_{2y}, \quad
    C_z^s = C_{1z} + C_{2z}, \quad C_z^a = C_{1z} - C_{2z}. 
    \label{eq:sym_def}
\end{equation}

Expressing the boundary conditions (\ref{eq:bc_Hy_plus})--(\ref{eq:bc_Ez_minus}) in terms of these new variables and keeping terms up to first order in $A d_P$, we obtain a significantly simplified system.

For the magnetic field:
\begin{subequations}
\begin{align}
    S_{1y} &= C_y^s + i A d_P C_y^a - \frac{\omega}{A^2} k_z \delta \tilde{p}_x, \label{eq:bc_Hy_plus_simp} \\
    S_{2y} &= C_y^s - i A d_P C_y^a - \frac{\omega}{A^2} k_z \delta \tilde{p}_x, \label{eq:bc_Hy_minus_simp} \\
    S_{1z} &= C_z^s + i A d_P C_z^a + \frac{\omega}{A^2} k_y \delta \tilde{p}_x, \label{eq:bc_Hz_plus_simp} \\
    S_{2z} &= C_z^s - i A d_P C_z^a + \frac{\omega}{A^2} k_y \delta \tilde{p}_x. \label{eq:bc_Hz_minus_simp}
\end{align}
\end{subequations}

For the electric field:
\begin{subequations}
\begin{align}
    \frac{1}{k_S} S_{1z} &= \frac{1}{k_P^2 A^2} \Bigg[ \alpha \Big( A (C_z^a + i A d_P C_z^s) - \omega \delta \tilde{p}_y \Big) + \frac{K_1}{2} \Big( A (C_y^a + i A d_P C_y^s) + \omega \delta \tilde{p}_z \Big) \Bigg], \label{eq:bc_Ey_plus_simp} \\
    \frac{1}{k_S} S_{1y} &= \frac{1}{k_P^2 A^2} \Bigg[ \frac{K_1}{2} \Big( A (C_z^a + i A d_P C_z^s) - \omega \delta \tilde{p}_y \Big) + \beta \Big( A (C_y^a + i A d_P C_y^s) + \omega \delta \tilde{p}_z \Big) \Bigg], \label{eq:bc_Ez_plus_simp} \\
    -\frac{1}{k_S} S_{2z} &= \frac{1}{k_P^2 A^2} \Bigg[ \alpha \Big( A (C_z^a - i A d_P C_z^s) - \omega \delta \tilde{p}_y \Big) + \frac{K_1}{2} \Big( A (C_y^a - i A d_P C_y^s) + \omega \delta \tilde{p}_z \Big) \Bigg], \label{eq:bc_Ey_minus_simp} \\
    -\frac{1}{k_S} S_{2y} &= \frac{1}{k_P^2 A^2} \Bigg[ \frac{K_1}{2} \Big( A (C_z^a - i A d_P C_z^s) - \omega \delta \tilde{p}_y \Big) + \beta \Big( A (C_y^a - i A d_P C_y^s) + \omega \delta \tilde{p}_z \Big) \Bigg]. \label{eq:bc_Ez_minus_simp}
\end{align}
\end{subequations}

We now eliminate the superconducting amplitudes $\{S_{1y}, S_{1z}, S_{2y}, S_{2z}\}$ from the simplified boundary conditions (\ref{eq:bc_Hy_plus_simp})--(\ref{eq:bc_Ez_minus_simp}). This yields a closed linear system for the four unknown coefficients $\hat{C} = (C_y^s, C_y^a, C_z^s, C_z^a)^T$ that describe the magnetic field profile in the ferroelectric layer:
\begin{equation}
    \hat{K} \, \hat{C} = \hat{P}.
    \label{eq:matrix_system}
\end{equation}
The $4 \times 4$ matrix $\hat{K}$ is defined as
\begin{equation}
    \hat{K} = 
    \begin{pmatrix}
        \dfrac{i K_1 d_P}{2 k_P^2} & \dfrac{K_1}{2 A k_P^2} & -\dfrac{1}{k_S}+\dfrac{i \alpha d_P}{k_P^2} & \dfrac{\alpha}{A k_P^2} - \dfrac{i A d_P}{k_S} \\[8pt]
        -\dfrac{1}{k_S}+\dfrac{i \beta d_P}{k_P^2} & \dfrac{\beta}{A k_P^2} - \dfrac{i A d_P}{k_S} & \dfrac{i K_1 d_P}{2 k_P^2} & \dfrac{K_1}{2 A k_P^2} \\[8pt]
        -\dfrac{i K_1 d_P}{2 k_P^2} & \dfrac{K_1}{2 A k_P^2} & \dfrac{1}{k_S}-\dfrac{i \alpha d_P}{k_P^2} & \dfrac{\alpha}{A k_P^2} - \dfrac{i A d_P}{k_S} \\[8pt]
        \dfrac{1}{k_S}-\dfrac{i \beta d_P}{k_P^2} & \dfrac{\beta}{A k_P^2} - \dfrac{i A d_P}{k_S} & -\dfrac{i K_1 d_P}{2 k_P^2} & \dfrac{K_1}{2 A k_P^2}
    \end{pmatrix}.
    \label{eq:K_matrix}
\end{equation}
To zeroth order in $A d_P$, the matrix $\hat{K}$ simplifies to
\begin{equation}
    \hat{K} = \left(
\begin{array}{cccc}
 \frac{i K_1 d}{2 k_P^2} & \frac{K_1}{2 A k_P^2} & -\frac{1}{k_S}+\frac{i \alpha  d}{k_P^2} & \frac{\alpha }{A k_P^2} \\
 -\frac{1}{k_S}+\frac{i \beta  d}{k_P^2} & \frac{\beta }{A k_P^2} & \frac{i K_1 d}{2 k_P^2} & \frac{K_1}{2 A k_P^2} \\
 -\frac{i K_1 d}{2 k_P^2} & \frac{K_1}{2 A k_P^2} & \frac{1}{k_S}-\frac{i \alpha  d}{k_P^2} & \frac{\alpha }{A k_P^2} \\
 \frac{1}{k_S}-\frac{i \beta  d}{k_P^2} & \frac{\beta }{A k_P^2} & -\frac{i K_1 d}{2 k_P^2} & \frac{K_1}{2 A k_P^2} \\
\end{array}
\right)
    \label{eq:K_matrix_0}
\end{equation}
The source term $\hat{P}$ is a vector proportional to the polarization fluctuation $\delta \tilde{\bm p} = (\tilde{p}_x, \tilde{p}_y, \tilde{p}_z)^T$:
\begin{equation}
    \hat{P} = \hat{P}_t \, \delta \tilde{\bm p}, \quad \text{where} \quad
    \hat{P}_t = \frac{\omega}{A^2}
    \begin{pmatrix}
        \dfrac{k_y}{k_S} & \dfrac{\alpha}{k_P^2} & -\dfrac{K_1}{2 k_P^2} \\[8pt]
        -\dfrac{k_z}{k_S} & \dfrac{K_1}{2 k_P^2} & -\dfrac{\beta}{k_P^2} \\[8pt]
        -\dfrac{k_y}{k_S} & \dfrac{\alpha}{k_P^2} & -\dfrac{K_1}{2 k_P^2} \\[8pt]
        \dfrac{k_z}{k_S} & \dfrac{K_1}{2 k_P^2} & -\dfrac{\beta}{k_P^2}
    \end{pmatrix}.
    \label{eq:Pt_matrix}
\end{equation}

\subsection{Electric Field in the Ferroelectric Layer}
The electric-field components within the ferroelectric can be expressed in terms of the symmetric/antisymmetric coefficients $\hat{C}$ and the polarization $\delta \tilde{\bm p}$. Using Eqs.~\eqref{eq:E_components} and \eqref{eq:E_in_P} we find
\begin{subequations}
\begin{align}
    \tilde{E}_x &= -\frac{1}{\omega \varepsilon_0} (C_z^s k_y - C_y^s k_z) - \frac{k_P^2}{A^2 \varepsilon_0} \tilde{P}_x, \label{eq:Ex_final} \\
    \tilde{E}_y &= \frac{\omega \mu_0}{k_P^2 A} \left( \alpha C_z^a + \frac{K_1}{2} C_y^a \right) + \frac{\omega^2 \mu_0}{k_P^2 A^2} \left( - \alpha \tilde{P}_y + \frac{K_1}{2} \tilde{P}_z \right), \label{eq:Ey_final} \\
    \tilde{E}_z &= -\frac{\omega \mu_0}{k_P^2 A} \left( \frac{K_1}{2} C_z^a + \beta C_y^a \right) - \frac{\omega^2 \mu_0}{k_P^2 A^2} \left( -\frac{K_1}{2} \tilde{P}_y + \beta \tilde{P}_z \right). \label{eq:Ez_final}
\end{align}
\end{subequations}

This result can be written compactly in matrix form as
\begin{equation}
    \tilde{\bm E} = \hat{\Gamma} \, \hat{C} + \hat{\Gamma}_P \, \delta \tilde{\bm p},
    \label{eq:E_matrix_form2}
\end{equation}
where the $3 \times 4$ matrix $\hat{\Gamma}$ and the $3 \times 3$ matrix $\hat{\Gamma}_P$ are given by
\begin{subequations}
\begin{align}
    \hat{\Gamma} &= 
    \begin{pmatrix}
        \dfrac{k_z}{\varepsilon_0 \omega} & 0 & -\dfrac{k_y}{\varepsilon_0 \omega} & 0 \\[8pt]
        0 & \dfrac{K_1 \mu_0 \omega}{2 A k_P^2} & 0 & \dfrac{\alpha \mu_0 \omega}{A k_P^2} \\[8pt]
        0 & -\dfrac{\beta \mu_0 \omega}{A k_P^2} & 0 & -\dfrac{K_1 \mu_0 \omega}{2 A k_P^2}
    \end{pmatrix}, \\
    \hat{\Gamma}_P &= 
    \begin{pmatrix}
        -\dfrac{k_P^2}{A^2 \varepsilon_0} & 0 & 0 \\[8pt]
        0 & -\dfrac{\alpha \mu_0 \omega^2}{A^2 k_P^2} & \dfrac{K_1 \mu_0 \omega^2}{2 A^2 k_P^2} \\[8pt]
        0 & \dfrac{K_1 \mu_0 \omega^2}{2 A^2 k_P^2} & -\dfrac{\beta \mu_0 \omega^2}{A^2 k_P^2}
    \end{pmatrix}.
    \label{eq:Gamma_matrices}
\end{align}
\end{subequations}

The linear response of the system, relating the induced depolarization electric field to the polarization fluctuation, is encoded in the response tensor $\hat{N}$, defined by $\tilde{\bm E} = - \hat{N} \, \delta \tilde{\bm p}$. Combining Eqs.~(\ref{eq:matrix_system}), (\ref{eq:E_matrix_form2}), and eliminating $\hat{C}$, we obtain
\begin{equation}
    \hat{N} = -\left( \hat{\Gamma} \, \hat{K}^{-1} \, \hat{P}_t + \hat{\Gamma}_P \right).
    \label{eq:N_definition}
\end{equation}

After evaluating this expression in the thin-film limit, the response tensor takes a remarkably simple diagonal form\begin{equation}
    \hat{N} = \operatorname{diag}\big( N(k, \omega),\, 0,\, 0 \big),
    \label{eq:N_diagonal}
\end{equation}
demonstrating that only the $x$-component of the polarization fluctuation ($\delta p_x$) generates a depolarization field inside the S/FE/S structure.

The non-zero element $N(k, \omega)$ is given by:
\begin{equation}
    N(k,\omega)=\frac{\omega^2}{\varepsilon_0[\omega^2-c^2k^2 d_P/(d_P + \lambda_{{\rm eff}})]},
    \label{eq:N_full}
\end{equation}
where $k^2 = k_y^2 + k_z^2$. Equation (\ref{eq:N_full}) is the central result of this supplemental calculation. It explicitly shows how the superconducting screening (via $\lambda_{\text{eff}}$) modifies the depolarization field compared to a standalone ferroelectric film.

\section{Quantization of polarization fluctuations}

The Hamiltonian of the system is composed of the kinetic and potential energies, which reads 
\begin{align}
    \hat{H}&=\int_{{\bf r}\in V_P} \left(\frac{1}{2}m_p|\partial_t{\bf P}({\bf r})|^2+{\cal F}({\bf r}) \right)d{\bf r}\approx \hat{H}_0+\hat{H}_2,
    \label{Ham_tot}
\end{align}
where $\hat{H}_0$ is the free energy in the absence of the fluctuation, $\hat{H}_{2}$ denote the terms of the square of the fluctuation $\delta {\bf p}$, $m_p=1/(\varepsilon_0\Omega_p^2)$ is the polarization inertia, and $V_P$ is the volume of the ferroelectric insulator.
The terms that are linear in $\delta p$ vanish automatically because the free energy is minimized.

The quadratic part of $\hat{H}$ reads
\begin{align}
    \hat{H}_2&=\int_{{\bf r}\in V_P} \left\{\frac{1}{2}m_p|\partial_t{\bf P}({\bf r},t)|^2+\frac{1}{2}(\alpha_1+3\alpha_2P^2_{0y})\delta p^2_y({\bf r},t)\right. \nonumber \\
    &+\left.\frac{\alpha_3}{2}\left[\delta p^2_x({\bf r},t)+\delta p^2_z({\bf r},t)\right]-\frac{1}{2}{\bf E}_d({\bf r},t)\cdot \delta {\bf p}({\bf r},t)\right\} d{\bf r},
    \label{H2}
\end{align}
in which the contribution of the static electric field to the dispersion is included in $P_{0y}$. 
The polarization fluctuation that is uniform across the ferroelectric film can be decomposed as a linear superposition of the eigenmodes expressed by Eq.~(17) of the main text:
\begin{align}
    \begin{pmatrix}
        \delta {p}_{x}({\bf r},t) \\
        \delta {p}_{y}({\bf r},t)\\
        \delta {p}_{z}({\bf r},t)
    \end{pmatrix}=\frac{1}{\sqrt{V_P}}\sum_{\lambda,{\bf k}}{Q}_{\lambda,{\bf k}}(t){\bf e}_{\lambda,{\bf k}}e^{i{\bf k}\cdot \pmb{\rho}}\left[\Theta(x+d_P)-\Theta(x-d_P)\right].
    \label{pk}
\end{align}
Since $\delta {\bf p}({\bf r},t)$ must be real, i.e., $\delta {\bf p}^{\ast}=\delta {\bf p}$, thus, the amplitude ${Q}_{\lambda,{\bf k}}$ satisfies
\begin{align}
    {Q}^{\ast}_{\lambda,{\bf k}}(t)=Q_{\lambda,-{\bf k}}(t).
    \label{Q_relation}
\end{align}
$Q_{\lambda,{\bf k}}$ is also referred to as the canonical coordinate, and the canonical momentum is defined as $\Pi_{\lambda,{\bf k}}=m_p\partial_t Q_{\lambda,{\bf k}}$.

We define the Fourier components $\delta {\bf p}_{\bf k}$ of $\delta {\bf p}({\bf r})$ according to 
\[
\delta {\bf p}({\bf r})=\frac{1}{\sqrt{V_P}}\sum_{\bf k}\delta {\bf p}_{\bf k}e^{i{\bf k}\cdot \pmb{\rho}}\left[\Theta(x+d_P)-\Theta(x-d_P)\right].
\]
According to Eq.~(\ref{pk}), 
\begin{align}
\delta {\bf p}_{\bf k}=\sum_{\lambda}{Q}_{\lambda,{\bf k}}(t){\bf e}_{\lambda,{\bf k}}.
\end{align}
Substituting into Eq.~(\ref{H2}), we obtain
\begin{align}
    \hat{H}_2=\sum_{\lambda,{\bf k}}\left(\frac{1}{2m_p}{\Pi}^{\ast}_{\lambda,{\bf k}}{\Pi}_{\lambda,{\bf k}}+\frac{1}{2}m_p\omega^2_{\lambda,{\bf k}} {Q}^{\ast}_{\lambda,{\bf k}} {Q}_{\lambda,{\bf k}}\right).
    \label{H0}
\end{align}

To quantize the system, we suppose the quantization  
\begin{align}
    \hat{Q}_{\lambda,{\bf k}}=C_{\lambda,{\bf k}}(\hat{a}_{\lambda,{\bf k}}+\hat{a}^{\dagger}_{\lambda,-{\bf k}})
\end{align}
in terms of the coefficients $C_{\lambda,{\bf k}}$ and the bosonic creation (annihilation) operator of ferrons $\hat{a}^{\dagger}_{\lambda,{\bf k}}$ $(\hat{a}_{\lambda,{\bf k}})$ for the $\lambda$-mode. Thereby, 
\begin{align}
    \hat{\Pi}_{\lambda,{\bf k}}=m_p\partial_t \hat{Q}_{\lambda,{\bf k}}=-im_p\omega_{\lambda,{\bf k}}C_{\lambda,{\bf k}}(\hat{a}_{\lambda,{\bf k}}-\hat{a}^{\dagger}_{\lambda,-{\bf k}}).
\end{align}
Substituting into Eq.~(\ref{H0}) and supposing  $\hat{H}_2=\sum_{\lambda,{\bf k}}\hbar\omega_{\lambda,{\bf k}}(\hat{a}^{\dagger}_{\lambda,{\bf k}}\hat{a}_{\lambda,{\bf k}}+1/2)$, we obtain
\begin{align}
    C_{\lambda,{\bf k}}=\sqrt{\frac{\hbar}{2m_p\omega_{\lambda,{\bf k}}}},
\end{align}
with which
\begin{align}
    \hat{Q}_{\lambda,{\bf k}}&=\sqrt{\frac{\hbar}{2m_p\omega_{\lambda,{\bf k}}}}(\hat{a}_{\lambda,{\bf k}}+\hat{a}^{\dagger}_{\lambda,-{\bf k}}),\nonumber\\
    \hat{\Pi}_{\lambda,{\bf k}}&=-i\sqrt{\frac{\hbar m_p\omega_{\lambda,{\bf k}}}{2}}(\hat{a}_{\lambda,{\bf k}}-\hat{a}^{\dagger}_{\lambda,-{\bf k}}),
    \label{Qk_final}
\end{align}
which are consistent with Eq.~(\ref{Q_relation}).
Accordingly, the quantized expression for the polarization fluctuation reads
\begin{align}
    \begin{pmatrix}
        \delta \hat{p}_{x}(\bf{r}) \\
        \delta \hat{p}_{y}(\bf{r})\\
        \delta \hat{p}_{z}(\bf{r})
    \end{pmatrix} =\frac{1}{\sqrt{V_P}}\sum_{\lambda,{\bf k}}\sqrt{\frac{\hbar}{2m_p\omega_{\lambda,{\bf k}}}}{\bf e}_{\lambda,{\bf k}}(\hat{a}_{\lambda,{\bf k}}+\hat{a}^{\dagger}_{\lambda,-{\bf k}})e^{i{\bf k}\cdot \pmb{\rho}}\left[\Theta(x+d_P)-\Theta(x-d_P)\right].
    \label{delta_p_a2}
\end{align}

\section{Quantization of the Swihart mode}

We then address the electromagnetic modes in the S/vacuum/S heterostructures, in which the thickness of the middle non-magnetic insulator is $2d_P$. Swihart found that the superconductor strongly modulates the electromagnetic field distribution, significantly slowing the speed of electromagnetic waves, a phenomenon now known as the Swihart mode \cite{Swihart1961}.

The first mode is the TE mode with the dispersion relation 
\begin{equation}
    \omega=\frac{1}{\sqrt{\mu_0\epsilon_0}}\sqrt{k^2+\frac{1}{2}\left(\frac{1}{\lambda_{\rm eff}^2}+\frac{1}{d_P\lambda_{\rm eff}}\right)},
\end{equation}
which turns out to be a high frequency and hence mismatches with the ferron modes. The second one is the TM mode with the Swihart-mode frequency  
\begin{equation}
\Omega_s({\bf k})=\sqrt{\frac{d_P}{d_P+\lambda_{\rm eff}}}\frac{1}{\sqrt{\mu_0\epsilon_0}}|{\bf k}|.
\label{Swihart_mode}
\end{equation}  

The magnetic field $\hat{\bf H}_{Sw}=\hat{H}_{Sw,y}\hat{\bf y}+\hat{H}_{Sw,z}\hat{\bf z}$ inside the interlayer for the Swihart photons quantized as \cite{Qiu2024}
\begin{align}
     \hat{H}_{{\rm Sw},y}&=\sum_{\bf k} \left(\frac{\Omega_s({\bf k})}{2k}\sqrt{\frac{\epsilon_I\hbar\Omega_s({\bf k})}{d_P S}}\cos\theta_{\bf k} e^{i{\bf k}\cdot{\pmb \rho}}\hat{p}_{\bf k}+{\rm H.c.}\right),\nonumber\\
     \hat{H}_{{\rm Sw},z}&=\sum_{\bf k} \left(-\frac{\Omega_s({\bf k})}{2k}\sqrt{\frac{\epsilon_I\hbar\Omega_s({\bf k})}{d_PS}}\sin\theta_{\bf k} e^{i{\bf k}\cdot{\pmb \rho}}\hat{p}_{\bf k}+{\rm H.c.}\right),
     \label{magnetic_field_quantization}
\end{align}
where $\theta_{\bf k}$ is the angle between the wave vector ${\bf k}$ and the $\hat{\bf z}$-direction and $S$ is the area of the S/FE interface. They are uniform across the insulator film. Then according to 
$\nabla\times {\bf H}=\partial_t{\bf D}=\epsilon_0\partial_t{\bf E}$, inside the interlayer, 
\begin{align}
    -i\epsilon_0\Omega_s({\bf k}){\bf E}&=\left(\frac{\partial H_z}{\partial y}-\frac{\partial H_y}{\partial z}\right)\hat{\bf x}-\frac{\partial H_z}{\partial x}\hat{\bf y}+\frac{\partial H_y}{\partial x}\hat{\bf z}=\left(\frac{\partial H_z}{\partial y}-\frac{\partial H_y}{\partial z}\right)\hat{\bf x}\nonumber\\
        &=-i\sum_{\bf k} \left(\frac{\Omega_s({\bf k})}{2}\sqrt{\frac{\epsilon_0\hbar\Omega_s({\bf k})}{d_PS}}e^{i{\bf k}\cdot{\pmb \rho}}\hat{p}_{\bf k}+{\rm H.c.}\right)\hat{\bf x}.
\end{align}
So the electric field is quantized according to
\begin{align}
    {\bf E}_{\rm Sw}({\bf r})=\sum_{\bf k} \frac{1}{2\epsilon_0}\sqrt{\frac{\epsilon_I\hbar\Omega_s({\bf k})}{d_PS}}e^{i{\bf k}\cdot{\pmb \rho}}\left(\hat{p}_{\bf k}+\hat{p}^{\dagger}_{-\bf k}\right)\hat{\bf x},
\end{align}
which is along the film normal $\hat{\bf x}$-direction.

\end{widetext}

\bibliography{S_FE_S}

\end{document}